# ESO IMAGING SURVEY:
## Past Activities and Future Prospects

*L. DA COSTA[1], S. ARNOUTS[1], C. BENOIST[1], E. DEUL[1,2], R. HOOK[3], Y. -S. KIM[1],*
*M. NONINO[1,4], E. PANCINO[1,5], R. RENGELINK[1,2], R. SLIJKHUIS[1], A. WICENEC[1],*
*S. ZAGGIA[1]*

[1]*European Southern Observatory, Garching b. München, Germany*
[2]*Leiden Observatory, Leiden, The Netherlands*
[3]*Space Telescope – European Coordinating Facility, Garching b. München, Germany*
[4]*Osservatorio Astronomico di Trieste, Italy*
[5]*Dipartimento di Astronomia, Università di Padova, Italy*

## 1. Introduction

The ESO Imaging Survey (EIS) project is an ongoing effort to carry out public imaging surveys in support of VLT programmes. Background information on the original and future goals of the programme can be found in Renzini & da Costa (1997) and in a companion article by Renzini & da Costa in this issue of *The Messenger*.

The first phase of the project, which started in July 1997, consisted of a moderately deep, large-area survey (EIS-WIDE) and a deep optical/infrared survey (EIS-DEEP), with the observations being conducted at the NTT. EIS has recently reached another milestone with the completion of a Pilot Survey using the Wide-Field Imager (WFI) mounted on the MPG/ESO 2.2-m telescope at La Silla.

The purpose of this contribution is to briefly review the results of the original EIS and to give an update of the results obtained from the observations carried out as part of the Pilot Survey. The ongoing work to develop an advanced pipeline for handling data from large CCD mosaics and of facilities to make access to the data products easier to external users are also discussed.

## 2. EIS-WIDE

This part of the survey consisted of a mosaic of overlapping EMMI-NTT frames ($9 \times 9$ arcmin) with each position on the sky being sampled twice for a total integration time of 300 sec. The observations were carried out in the period July 97–March 98 and covered four patches of the sky south of $\delta = -20°$ and in the right ascension range $22^h < \alpha < 10^h$, thus producing targets suitable for the VLT almost year-round. The locations of these fields are shown in Figure 1 and their approximate centres are given in Table 1, which also lists the area covered in each passband. The typical depth of the survey, as measured from the co-added mosaics and with the magnitudes expressed in the AB system, is given in Table 2 which lists: in column (1) the passband; in column (2) the median seeing; in column (3) the $1\sigma$ limiting isophote; in column (4) the 80% completeness magnitude (e.g., Nonino et al. 1999); in column (5) the 3σ limiting magnitude within an aperture twice the median FWHM; and in column (6) the originally proposed magnitude limits (Renzini & da Costa 1997) expressed as in column (5). Note that these numbers are representative values since, as pointed out in the original papers (Nonino et al. 1999, Prandoni et al. 1999, Benoist et al. 1999), the observing conditions varied significantly during the course of the observations. Comparison of Tables 1 and 2 to the original goals of the survey (Renzini & da Costa 1997, and revised in da Costa 1997) shows that, while the desired limiting magnitudes were reached, not all the objectives were met. Among the most important omissions were the observations of a comparable area in V and of patch B in the U-band. The incompleteness was due to a combination of poor weather conditions and unexpected overheads in the observations. A total of about 42 nights were assigned to this part of the project out of which about 30% were lost due to the weather alone, compromising especially the coverage of EIS patches A and B. At the time, the Working Group (WG) in charge of overseeing the survey decided to postpone the U-band observations of patch B and the V-band observations of the other patches, giving priority to the I-band survey.

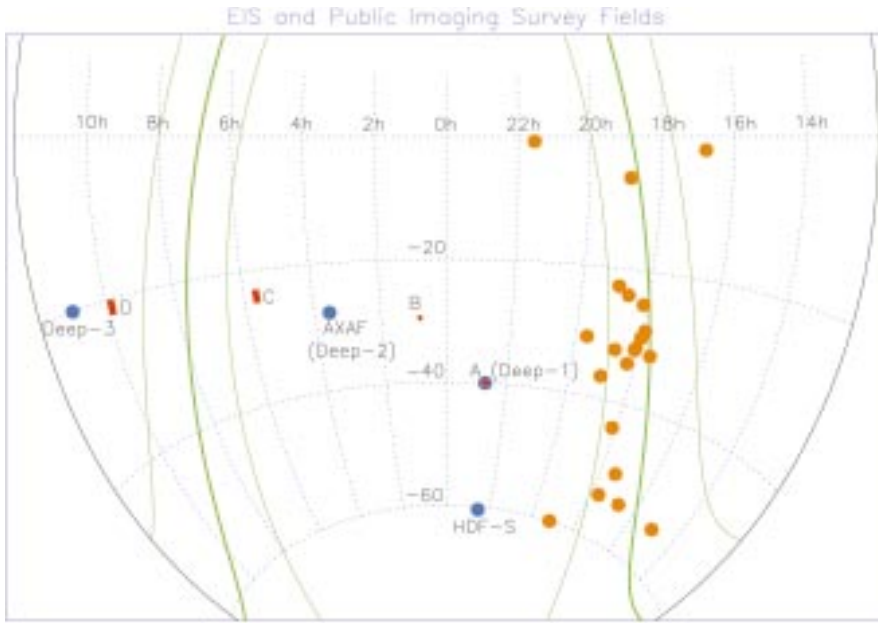

*Figure 1: Sky projection showing the location of the original EIS-WIDE patches (red), the EIS-DEEP AXAF and HDF-S fields (blue), and the stellar fields of the Pilot Survey (orange). Also in blue are the new fields selected for the Deep Public Survey to be carried out in periods 64–67. Note that one of the regions selected coincides with patch A and the other with the EIS-DEEP AXAF field (see Fig. 2). The third field (Deep-3) has been selected in a region with lower absorption and less crowding than patch D. The right ascension range was chosen to distribute as much as possible the observations throughout the year.*

*Table 1: EIS-WIDE: Sky Coverage (square degrees).*

| Patch | $\alpha_{2000}$ | $\delta_{2000}$ | B | V | I |
|-------|-----------------|-----------------|-----|-----|------|
| A     | 22 43 03        | –39 58 30       | –   | 1.2 | 3.2  |
| B     | 00 48 22        | –29 31 48       | 1.5 | 1.5 | 1.6  |
| C     | 05 38 24        | –23 51 54       | –   | –   | 6.0  |
| D     | 09 51 40        | –21 00 00       | –   | –   | 6.0  |
| Total |                 |                 | 1.5 | 2.7 | 16.8 |

Despite the smaller total area and the varying quality of the data, it was possible to meet most of the primary science goals of the survey. Particularly successful has been the *I*-band survey covering ~ 17 square degrees to a limiting magnitude $I_{AB}$ ~ 23.5, currently the largest survey of its kind in the southern hemisphere. In addition the survey provided ~ 3 square degrees in *V I* and ~ 1.5 square degrees in *B V I*. From these data, samples of distant clusters of galaxies and quasar candidates were compiled (Olsen et al. l999a,b, Scodeggio et al. 1999, Zaggia et al. 1999) and used in several follow-up observations by different teams. The data, comprising astrometrically and photometrically calibrated pixel maps, derived catalogues and target lists, were made publicly available in March and July 1998 (Nonino et al. 1999, Prandoni et al. 1999, Benoist et al. 1999), meeting the planned deadlines. To ease the access to the data, a web interface was created to enable external users to request survey products, to extract image cutouts and to examine the selected targets on-line.

The accuracy of the relative astrometry is ~ 0.03 arcsec, making the coordinates sufficiently accurate for the preparation of multi-object spectroscopic observations. The internal accuracy of the photometric zero-point has been estimated to be better than 0.1 mag, consistent with the results obtained from comparisons with other data (e.g., DENIS). However, as more data in different passbands became available, gradients of the photometric zero-point, on a degree scale, have been noticed and are currently being investigated. The results of these investigations will be reported in due time.

## 3. EIS-DEEP

This part of the survey consisted of deep optical/infrared observations using SUSI2 and SOFI at the NTT of the HDF-S and AXAF fields shown in Figure 1. The original observations were carried out in the period August–November 1998, with all the data being publicly released December 10–12, 1998 (da Costa et al. 1999b, Rengelink et al. 1999). The release included fully processed and calibrated images, single-passband and colour catalogues, and a tentative list of *U*-and *B* dropouts. The centres of the different fields observed with SUSI2 and SOFI, chosen to provide some overlap, are given in Table 3. The data available for these fields are summarised in Tables 4 and 5 which provide, for each field, the following information: in column (1) the passbands available; in column (2) the total integration time; in column (3) the number of frames; in column (4) the final FWHM as measured on the co-added image; in column (5) the 1σ limiting isophote; in col-umn (6) the limiting magnitude of a 3σ detection within an aperture twice the FWHM. In Table 4 the last column lists the originally proposed limiting magnitudes for EIS-DEEP, expressed as those in column (7), assuming a 1 arcsec seeing (da Costa & Renzini 1998). To help visualise the fields and the different sets of data available, Figure 2 shows a schematic view of the EIS-DEEP observations of the HDF-S and AXAF fields, with the colours indicating the completeness of the multi-colour data. In the case of HDF-S the location of the WFPC2, STIS and NIC3 HST fields and their relation to the notation of the EIS-DEEP fields are also shown.

*Table 2: EIS-WIDE: Limiting Magnitudes (AB system).*

| Filter | FWHM (arcsec) | $\mu_{lim}$ (mag arcsec$^{-2}$) | 80% | 3σ (2 × FWHM) | 3σ (proposed) (2 × FWHM) |
|--------|---------------|--------------------------------|------|---------------|--------------------------|
| B      | 1.0           | 26.5                           | 24.6 | 24.2          | 24.2                     |
| V      | 0.8           | 26.2                           | 24.4 | 24.0          | 24.5                     |
| I      | 0.8           | 25.5                           | 23.7 | 23.7          | 24.0                     |

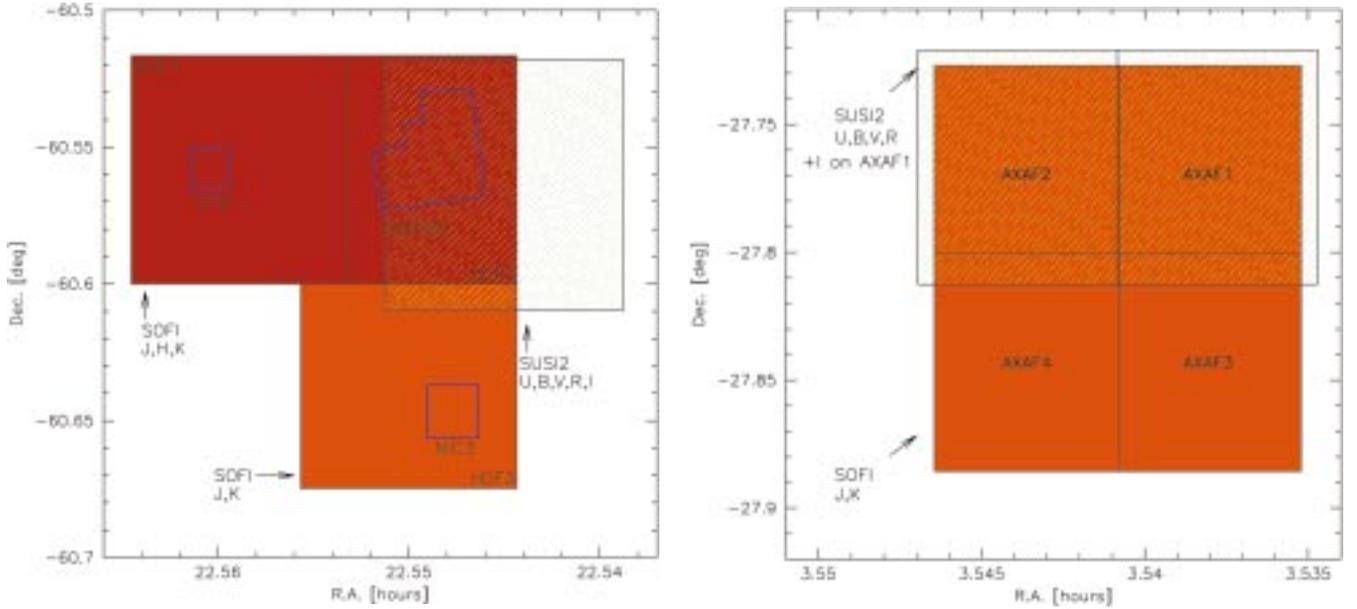

*Figure 2: Schematic view of the various data sets available for the HDF-S and AXAF fields from optical/infrared observations conducted with SUSI2 and SOFI at NTT. The different colours denote the completeness of the multi-colour data as indicated in the figure.*

Table 3: EIS-DEEP: SUSI2 and SOFI Pointings (J2000.0).

| Field | $\alpha_{SUSI2}$ | $\delta_{SUSI2}$ | $\alpha_{SOFI}$ | $\delta_{SOFI}$ |
|---|---|---|---|---|
| HDF1 | 22 33 29 | –60 33 50 | 22 33 32 | –60 33 30 |
| HDF2 | 22 32 42 | –60 33 50 | 22 33 00 | –60 33 30 |
| HDF3 | – | – | 22 33 00 | –60 37 59 |
| AXAF1 | 03 32 16 | –27 46 00 | 03 32 17 | –27 46 10 |
| AXAF2 | 03 32 38 | –27 46 00 | 03 32 37 | –27 46 10 |
| AXAF3 | – | – | 03 32 17 | –27 50 35 |
| AXAF4 | – | – | 03 32 37 | –27 50 35 |

Table 4: EIS-DEEP: HDF-S Observations (magnitudes in the AB system).

| Filter | $t_{total}$ (sec) | $N_f$ | FWHM (arcsec) | $\mu_{lim}$ (mag arcsec$^{-2}$) | 3σ (2 × FWHM) | 3σ (proposed) (2 × FWHM) |
|---|---|---|---|---|---|---|
| HDF1 | | | | | | |
| J | 10800 | 180 | 1.37 | 25.7 | 23.6 | 24.2 |
| H | 3600 | 60 | 0.91 | 25.1 | 23.4 | 23.7 |
| Ks | 10800 | 180 | 0.90 | 24.9 | 23.1 | 23.1 |
| HDF2 | | | | | | |
| U | 17800 | 22 | 1.00 | 28.2 | 26.3 | 26.6 |
| B | 6600 | 22 | 0.84 | 28.2 | 26.5 | 25.7 |
| V | 12250 | 49 | 1.27 | 28.3 | 26.5 | 25.8 |
| R | 5500 | 22 | 1.05 | 27.3 | 25.4 | 26.0 |
| I | 8800 | 44 | 1.11 | 27.0 | 25.0 | 26.2 |
| J | 10800 | 180 | 0.90 | 26.2 | 24.4 | 24.7 |
| H | 7200 | 120 | 0.85 | 24.9 | 23.2 | 24.2 |
| Ks | 18000 | 300 | 0.96 | 25.5 | 23.7 | 23.6 |
| HDF3 | | | | | | |
| J | 5820 | 97 | 1.18 | 25.5 | 23.5 | 24.2 |
| Ks | 7440 | 124 | 0.86 | 24.9 | 23.2 | 23.1 |

The highest priority was given to the observations of the HDF2 (WFPC2) field for which data in eight passbands are available, by and large reaching the required magnitude limits, with the notable exception of the *I*-band. Note that in this field the *J* and *Ks* observations reach half-magnitude fainter limits than those in the other fields, as suggested by the WG (da Costa & Renzini 1998). Even though it was not possible to complete the observations in all the passbands for all the fields, the data gathered provide ~ 150 square arcmin in *J Ks* and ~ 80 square arcmin in *UBVRI*. When combined, the optical/infrared deep observations provide a coverage of 15 and 25 square arcmin in eight and seven passbands, respectively. These numbers include the recently released data for the HDF-S STIS and NICMOS fields from observations conducted in July 1999 (see EIS home page). Unfortunately, due to bad weather, no new optical data were obtained during the four half nights scheduled in July 1999 at the NTT, and the *R I* images of HDF1, taken in 1998, proved to be unsuitable due to poor seeing and stray light. The incompleteness of the EIS-DEEP data relative to the original goals is primarily due to the optical observations which were affected by both poor weather and a variety of technical problems in 1998. An attempt to complete the optical coverage of the HDF1 (STIS) field will be made in June–July 2000, while the AXAF field has been chosen to be one of the three fields selected for the new Deep Public Survey as shown in Figure 1 (see Renzini & da Costa 1999).

## 4. WFI@2.2 Pilot Survey

With the announcement of the commissioning of the WFI@2.2m, the WG recommended the undertaking of a pilot multi-colour survey over the area already covered in $I$ – band, including the missing U-band observations of patch B. The motivation was to complete the original goals of EIS-WIDE and to steer the upgrade of the pipeline to cope with CCD-mosaics and the corresponding order of magnitude increase in the volume of data.

Even though originally proposed to be carried out in dark time over the observing Periods 62 and 63, some adjustments had to be made to accommodate the WFI commissioning period. In the end, a total of 30 hours was used for the Pilot Survey observations during the commissioning period in January 1999, and a total of 8.5 nights in the months of April, May and September. The observations consisted of five 4-minute dithered exposures, which allowed the coverage of 0.5 square degrees/hour in a single passband including overheads. A total of 18 square degrees were observed, covering patches C and D in $B$ and $BV$, respectively. The $B$-filter was used first because it was the only broad-band filter available during commissioning. In mid-September new observations of Patches A and B were conducted in $V$ and $U$ but the data have not yet been reduced at the time of writing. Table 6 shows the current status of the EIS-WIDE patches after combining the data available from EMMI at the NTT with those from the Pilot Survey, including the September 1999 run. These data together with the 17 hours scheduled in service mode for Period 64 should allow the completion of about 1.5 square degrees in $UBVI$ and 15 square degrees in $BVI$. These data should greatly broaden the range of applications that can benefit from the EIS public survey to produce samples for follow-up observations with the VLT. Preliminary estimates of the median seeing and limiting magnitudes reached in $B$ and $V$ from the data gathered so far using the WFI are given in Table 7.

Some of the scheduled time was unsuitable for the observation of the patches, either because the EIS-WIDE fields were not visible or because the nights were too bright. In these periods, a set of selected stellar fields were observed as a test case for future preparatory programmes for GIRAFFE and UVES using the fibre-positioner FLAMES, thus anticipating part of the Public Survey observations now planned for P64–67, the so-called Pre-Flames Survey recommended by the WG in their meeting in March 1999 (see Renzini & da Costa 1999). The selected fields (Figure 1) included Baade's window, globular (GC) and open clusters (OC). A total of 22 fields (5.5 square degrees) given in Table 8 were observed in different passbands. A full description of the available data can be found on the web.

Table 6: WFI Pilot Survey: Sky Coverage (square degrees).

| Patch | U | B | V | I |
|---|---|---|---|---|
| A | – | 1.0 | 1.2 | 3.2 |
| B | 1.25 | 1.5 | 1.5 | 1.6 |
| C | – | 6.0 | 6.0 | 6.0 |
| D | – | 6.0 | 6.0 | 6.0 |
| Total | 1.25 | 14.5 | 8.7 | 16.8 |

Table 7: WFI Pilot Survey: Limiting Magnitudes (AB system).

| Band | FWHM (arcsec) | $\mu_{lim}$ (mag arcsec$^{-2}$) | 80% | 3σ (2 × FWHM) |
|---|---|---|---|---|
| B | 0.9 | 26.8 | 24.5 | 25.0 |
| V | 1.0 | 26.4 | 24.1 | 24.6 |

Table 5: EIS:DEEP: AXAF Observations (magnitudes in the AB system).

| Filter | $t_{total}$ (sec) | $N_f$ | FWHM (arcsec) | $\mu_{lim}$ (mag arcsec$^{-2}$) | 3σ (2 × FWHM) |
|---|---|---|---|---|---|
| AXAF1 | | | | | |
| U | 17000 | 21 | 0.90 | 28.0 | 26.1 |
| B | 6600 | 22 | 1.10 | 28.0 | 26.4 |
| V | 5500 | 22 | 0.88 | 27.7 | 25.7 |
| R | 5500 | 22 | 0.89 | 27.5 | 25.9 |
| I | 12600 | 21 | 1.31 | 27.0 | 24.9 |
| J | 10800 | 180 | 0.99 | 25.7 | 23.8 |
| Ks | 10800 | 180 | 0.95 | 24.9 | 23.2 |
| AXAF2 | | | | | |
| U | 13000 | 16 | 0.90 | 27.8 | 26.0 |
| B | 5400 | 18 | 1.10 | 27.9 | 25.8 |
| V | 5500 | 22 | 0.88 | 27.7 | 26.1 |
| R | 5500 | 22 | 0.89 | 27.5 | 25.7 |
| J | 10800 | 180 | 0.90 | 25.8 | 24.1 |
| Ks | 10800 | 180 | 0.90 | 24.8 | 23.1 |
| AXAF3 | | | | | |
| J | 10800 | 180 | 0.90 | 25.4 | 23.7 |
| Ks | 10800 | 180 | 1.00 | 24.7 | 22.9 |
| AXAF4 | | | | | |
| J | 10800 | 180 | 0.70 | 25.6 | 24.1 |
| Ks | 7200 | 120 | 1.00 | 24.4 | 22.6 |

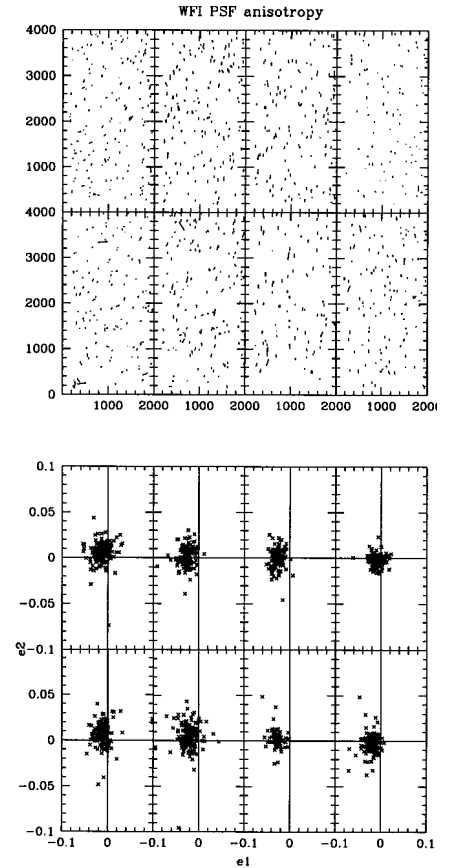

Figure 3: Vector representation of the pattern of the PSF anisotropy and the amplitude of its components for each chip of the CCD mosaic, as measured on a V-band image obtained from a 20-min exposure.

Figure 4: Colour image of a section of EIS patch D combining BV images of WFI with I-images of EMMI. ▶

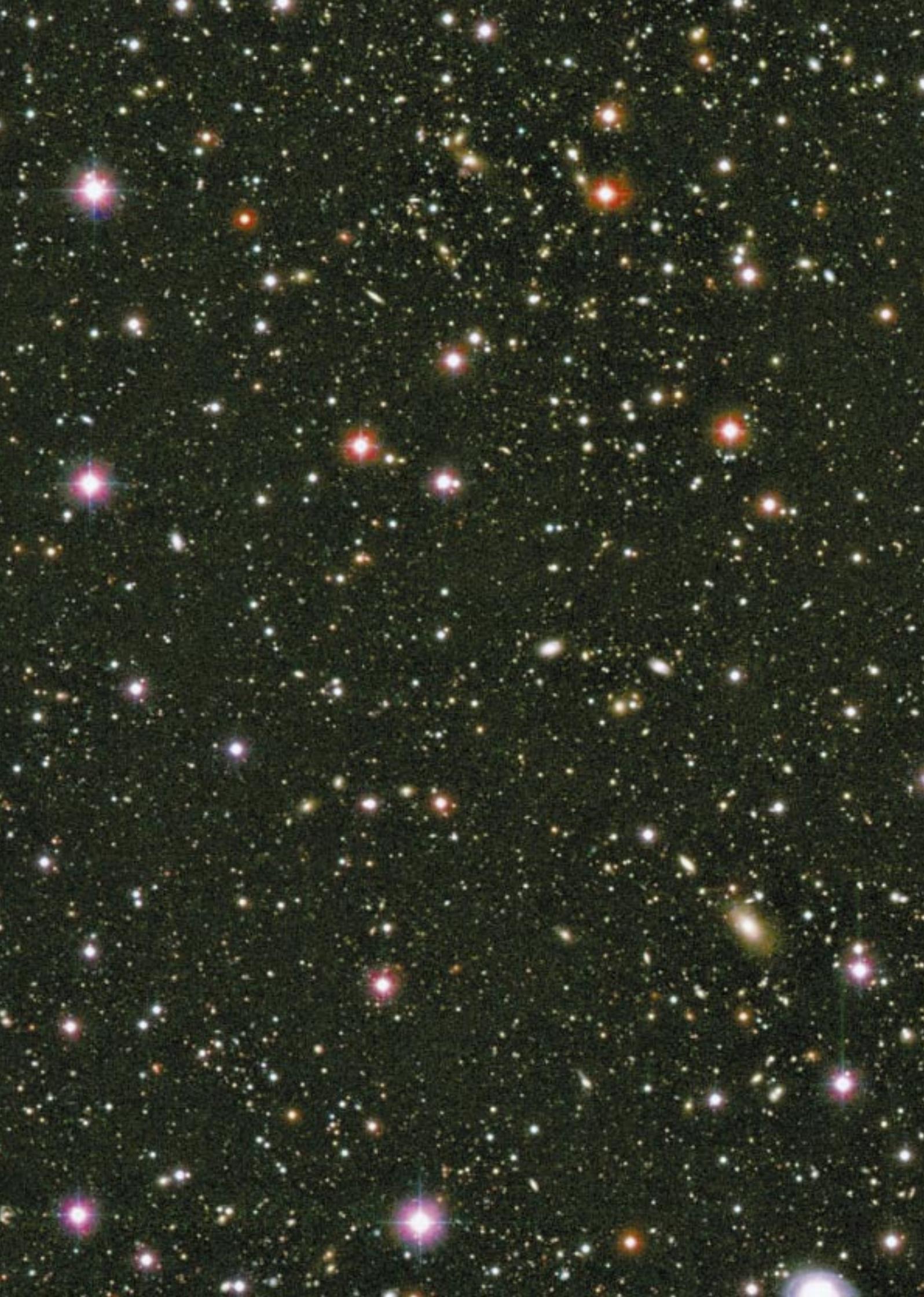

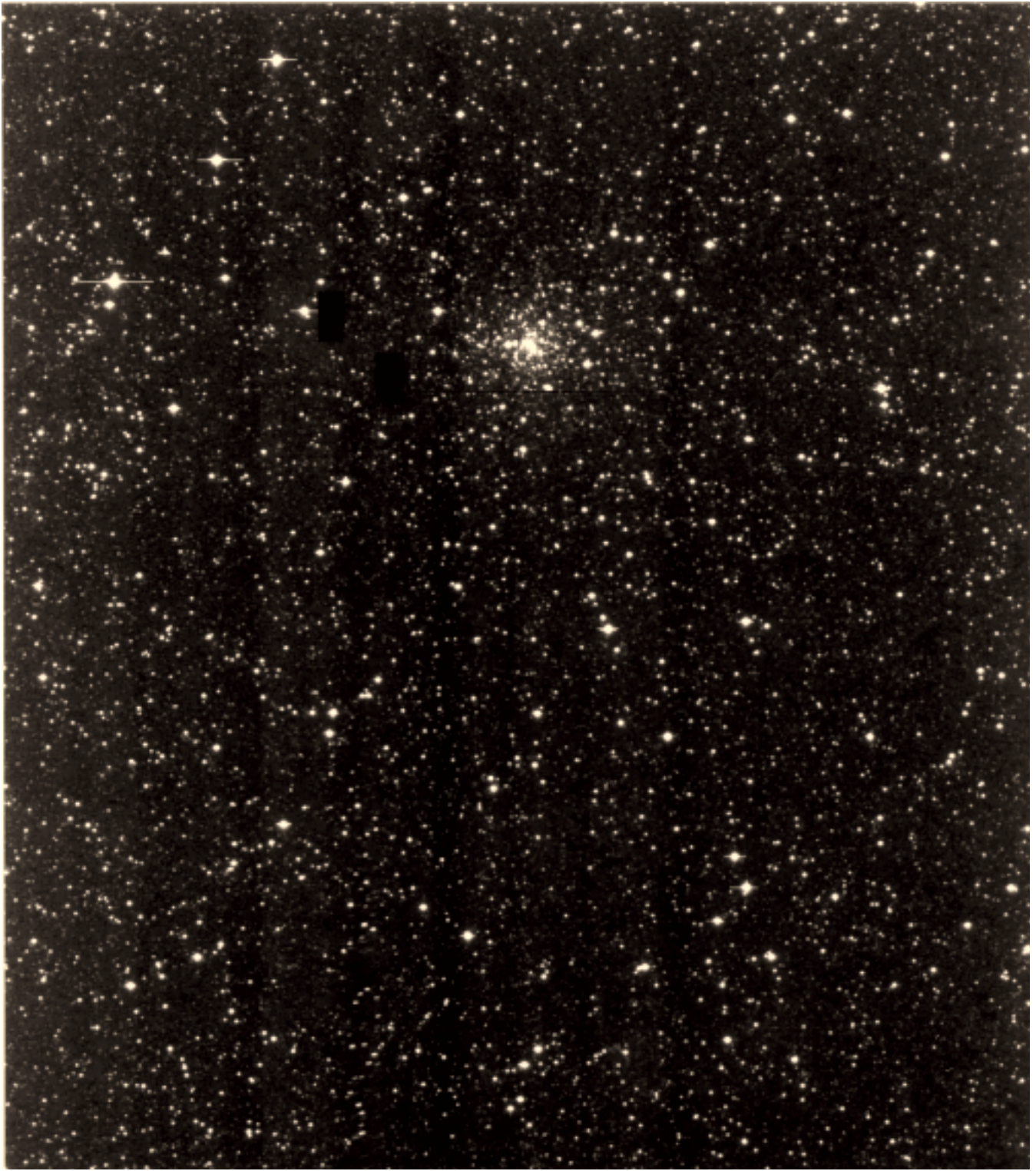

Figure 5: A 16.1 × 9.4 arcmin section of the co-added I-band image, rebinned to a scale of 0.5 arcsec/pixel, of the cluster NGC 6558 located in Baade's window. The regions devoid of objects are associated with the interchip gaps.

Altogether the observations from January to April 1999 yielded 400 Gby of raw data at a rate of about 40 Gby/night (0.5 Mb/sec), including calibration frames (standard stars, dome and sky flats). All the data have been processed, using the upgraded EIS-WFI pipeline, and publicly released on September 9, 1999.

Preliminary results show that the performance of the WFI@2.2 exceeds expectations and demonstrate its competitiveness in the area of wide-field imaging surveys. Analysis of the images shows that the point-spread function is very uniform across the entire field of view and that PSF distortions are small ~ 2%. This is illustrated in Figure 3 where the spatial distribution of the polarisation vector, representing the amplitude and direction of the PSF distortions, and the distribution of the amplitude of its two components are shown for each CCD chip of the WFI camera.

To illustrate the results obtained by the pipeline, Figure 4 shows a colour composite of a section of patch D (11 × 19 arcmin) constructed from the combination of WFI *B* and *V* mosaics with the EMMI *I*-band mosaic. Since one of the primary goals of Pilot Survey is to complement the *I*-band observations, the co-addition of the WFI images was carried out on the EMMI pixel scale, about 10% larger than that of the WFI. By dithering the WFI exposures and by co-adding the available frames using the weight maps, the final coadded mosaic shows no noticeable evidence of the inter-chip gaps, with the least sampled regions having an exposure time about 60% of that obtained in the best sampled regions. Another example can be found in Figure 5, where the co-added *B*-band image of the globular cluster NGC 6558, observed with 0.6 arcsec seeing, is shown. Note that in this case only two images have been co-added and the regions devoid of stars are the remains of the interchip gaps.

The co-addition is carried out after the astrometric calibration of each input frame. Figure 6 shows the distribution of the differences in the coordinates of 700,000 pairs of stars in patch D from independent astrometric solutions obtained for different WFI *V*-band frames. The rms of the distribution is ~ 0.04 arcsec, an estimate of the internal accuracy of the astrometric solution. This result shows that the position of the targets extracted from the survey data can be directly used to position the slits and fibers of VLT spectrographs such as FORS1, FORS2, VIMOS, NIRMOS and FLAMES. However, even though a satisfactory astrometric solution was determined for the stellar field shown in Figure 5, problems are still found in calibrating crowded fields. Procedures to astrometrically calibrate these fields, including proper modelling of the optical distortions, are still being investigated. These issues as well as a full description of the observations, data reduction and derived products will be presented in forthcoming papers.

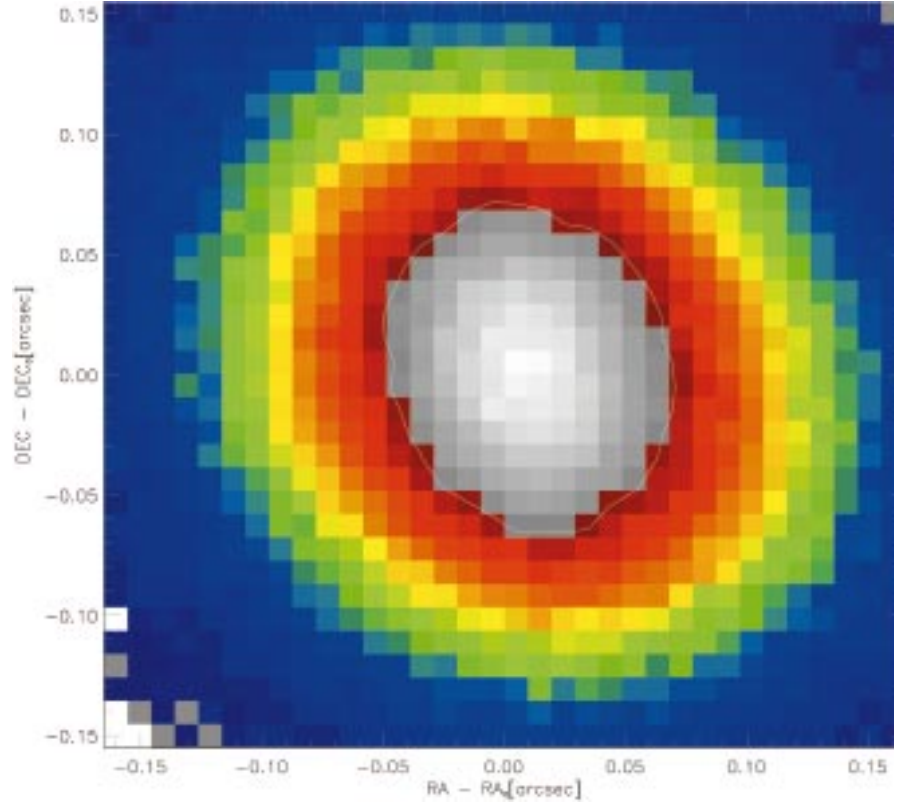

Figure 6: Distribution of the differences in the coordinates of 700,000 pairs of stars in patch D obtained from the independent astrometric calibration of different WFI frames. The solid contour corresponds to the half-maximum of the distribution, encompassing 62% of the points.

## 5. Data Handling

One of the most demanding tasks of the EIS project has been the development of a data reduction pipeline in parallel to the observations which have used four different instruments (EMMI, SUSI2, SOFI and WFI@2.2m) in a period of two years. Adaptation of the software to this rapidly changing environment has been by far the most challenging aspect of the project, which will hopefully subside in the next couple of years. This not only impacts the development effort but also the data reduction and verification which depend on the stability of the pipeline.

The original EIS pipeline was developed over a period of one year, starting in April 1997, by assembling existing software (IRAF, SExtractor, LDAC and Drizzle) adapted to the specific needs (detector and observing strategy) of the EIS-WIDE survey. The task was facilitated because most tools were devel-oped by astronomers in the Member States who participated in this development. At the time, the data volume was relatively small, which greatly facilitated the handling and processing of the data. Most of the work focused on the interfacing of the different processing blocks and in producing scripts to control the processing essentially in batch mode. A lot of effort was also spent in the design and implementation of a database which was primarily used for bookkeeping.

Table 8: WFI Pilot Survey: Stellar Fields.

| Object Name | Object Type | $\alpha_{1950}$ | $\delta_{1950}$ | Filters |
|---|---|---|---|---|
| Baade Window (1) | Bulge | 17 55 28 | –28 45 20 | B, V, I, z |
| Baade Window (2) | Bulge | 17 56 04 | –29 13 14 | B, V, I, z |
| Baade Window (3) | Bulge | 18 00 00 | –29 52 13 | B, V, I |
| Baade Window (4) | Bulge | 18 05 33 | –31 27 04 | B, V, I |
| Baade Window (5) | Bulge | 18 07 02 | –31 46 27 | V, I |
| Baade Window (6) | Bulge | 18 15 10 | –33 58 38 | V, I |
| NGC 6218 | GC | 16 47 14 | –01 56 52 | B, V, I |
| NGC 6397 | GC | 17 40 41 | –53 40 25 | B, V, I |
| NGC 6544 | GC | 18 07 20 | –24 59 51 | B. V, I |
| NGC 6809 | GC | 19 39 59 | –30 57 44 | B, V, I |
| NGC 7089 | GC | 21 33 29 | –00 49 23 | B, V, I |
| NGC 6752 | GC | 19 10 51 | –59 58 55 | B, V, I |
| NGC 6541 | GC | 18 08 02 | –43 42 20 | B, V, I |
| NGC 6681 | GC | 18 43 12 | –32 17 31 | B, V, I |
| NGC 6656 | GC | 18 36 24 | –23 54 12 | B, V, I |
| NGC 6717 | GC | 18 55 06 | –22 42 03 | B, V, I |
| NGC 6723 | GC | 18 59 33 | –36 37 54 | B, V, I |
| IC 4651 | OC | 17 24 42 | –49 57 00 | B, V, I |
| NGC 6405 | OC | 17 36 48 | –32 11 00 | B, V, I |
| NGC 6208 | OC | 16 45 30 | –53 4400 | B, V, I |
| NGC 5822 | OC | 15 01 30 | –54 09 00 | B, V, I |
| NGC 6705 | OC | 18 51 06 | –06 16 00 | B, V, I |

After the first EIS data release in March 1998, an upgrade was carried out to incorporate ECLIPSE (Devillard, Jung & Cuby 1999) for the reduction of SOFI data and to handle the two-CCD mosaic of SUSI2. Even though time-consuming, implementing these add-ons did not require major changes in the design of the pipeline.

The advent of the WFI, on the other hand, required a major overhaul driven by the sheer volume of the data. New hardware was required as well as a major re-design of the pipeline architecture. Over the past year new machines have been added, the number of processors available in each server increased, disk and memory capacities expanded and DLT tape units installed. In parallel, during the past six months significant modifications have been made to the pipeline software, even though the basic algorithms remain essentially the same. A more sophisticated control system has been developed to administer the resources of the system, to retrieve/store images in the tape library, to supervise the reduction process and to communicate with an expanded implementation of the database in a variety of ways. The new pipeline allows the reduction of both single CCD (SOFI and other single-chip instruments) and mosaics (e.g., SUSI2, WFI) and parts of the code have been optimised and parallelised to increase the overall throughput to match the expected input data rate. Finally, the pipeline has been made more user-friendly and uniform in order to facilitate its maintenance and integrity, to prevent discontinuities caused by the constant changes of Survey Team members and to allow it to be run by other interested users.

Broadly speaking the pipeline consists of four logical blocks:

1. Image Processing – this module performs the usual image processing steps removing instrument signatures and astrometrically and photometrically calibrating each individual WFI frame. It then performs the co-addition of dithered WFI frames, producing a single FITS image from which object catalogues are extracted. These are the basic products produced automatically by the pipeline which are then transferred to the archive for quick-look releases.

2. Image Mosaicing – this module deals with the production of image mosaics, which implies adjustments of the photometric zero-point, of final object catalogues, extracted from the co-added images using the most recent and extensively tested version of SExtractor, and of colour catalogues constructed either by the association of objects detected in each passband or by the use of a reference image (e.g., $\chi^2$-image).

3. Quality control – this module carries out the quality control of the data. It monitors image distortions, computes the rms distribution of the residuals of the astrometric solution, compares the astrometry of the objects extracted in different passbands and checks the spatial uniformity of the photometric zero-point. It also performs a preliminary scientific evaluation of the data. It culls object catalogues, computes simple statistics such as number counts, colour-colour diagrams, colour distributions and the two-point angular correlation function for objects classified as galaxies. The results are compared with other available data to check the photometric zero-points, the star-galaxy classification and uniformity of the object catalogues.

4. Target Selection – this module is used to produce a variety of target lists. It also produces image postage stamps (typically $2 \times 2$ arcmin) in different passbands of the selected objects for visual inspection. In time, other facilities are expected to be added, in particular for searches for variable and proper-motion objects. Among many possible applications, a major goal is to define secondary-standards in the surroundings of the traditional Landolt fields, essential to monitor the zero-point of the different chips in the CCD-mosaic.

Even though the backbone of the pipeline is in place, a lot of work remains to be done, especially on the control system, in order to allow the process to be fully automated. Time benchmarks of different parts of the pipeline demonstrate that most of the modules are operating at a rate ~ 0.2 Mpixels/sec. For instance, it currently takes about 7 hours to process 20 WFI frames and produce the results required for a quick-look release of a square degree worth of data. This corresponds to 2 hours of observations in the wide-angle-survey mode adopted by the Pilot Survey. Even though preliminary, these numbers show that the current pipeline is capable of coping with the volume of data being produced by the survey. Furthermore, the pipeline is not yet fully implemented and tested and there still is considerable room for improving the system throughput. The final consolidation of the software should last until the end of 1999, with a final target date of April 1, 2000. A detailed description of the available software will be presented in the "EIS-WFI Pipeline Primer" currently in preparation by the Survey Team.

## 6. Data Access

Unlike radio surveys and other public surveys carried out in space, ground-based optical imaging surveys require considerably more information to characterise the data available. An arduous and time-consuming task, given the time pressure and limited personnel, has been to establish adequate channels of communication between the Survey Team and the community at large to describe the progress of the project and to provide the necessary information about the observations and the various data products available. In this first phase this was accomplished by posting as much information as possible on the web, by periodically reviewing the status of the project in *The Messenger* (da Costa 1997, da Costa et al. 1998, da Costa & Renzini 1998) and by submitting and circulating papers with each data release.

During the first phase, requests for large volumes of data from different groups have been processed by the ESO Archive Group and about 100 Gby of data have been delivered in various forms. The requests originated from 11 countries, including all of the ESO member states. According to current guidelines, catalogues are available world-wide, but images are restricted to ESO member states, with the exception of those of the HDF-S fields. Table 9 shows a summary of data requests listing the product and the number of requests up to September 1999. The access to the EIS home page has increased by a factor of 5 over the period of the survey. It currently averages about 30,000 hits/month, reaching peaks twice as high after a data release. About one-third of these hits are from research institutes in Europe and abroad, constantly monitoring the progress of the survey, while the rest are from the general public. In this respect, it is worth noting the popularity of the EIS image gallery, with about 35,000 images being retrieved from the web. A total of about 4500 data requests have been processed so far, about half of which asked for derived products such as target lists and catalogues.

*Table 9: Summary of Data Requests (September 1999).*

| Product | Number of Requests |
|---|---|
| Bulk Data | 50 |
| Cutouts | 1128 |
| Mini-images | 319 |
| Target Lists | 737 |
| EIS-Deep catalogues | 1440 |
| HDF-S images | 821 |
| TOTAL | 4495 |
| Gallery Images | 34538 |

By expanding the scope of the observations to other areas of application and by improving the interface with the public, these numbers are likely to increase significantly.

## 7. Future Developments

With the recent approval of a two-year Large Programme to conduct a Public Survey (see Renzini & da Costa 1999) with the WFI and SOFI, it is now possible to envision a more stable operation leading to the production of more standard data sets. In this framework, new ways of making the data accessible to the community will be explored. Quick-look releases of the data, distribution of compressed data and the implementation of new on-line features are being considered as described by Renzini & da Costa (1999). These, together with the other products already available should provide users with a large range of options and greater flexibility to explore the available data.

A considerable effort is also underway to implement a flexible design to store the objects extracted from the public survey images in a database which will allow users to directly select and display objects according to their positions in the sky, their magnitudes and their colours without any reference to a specific catalogue. The database will be a dynamic entity that will grow in time as information from new observations of different sky regions or repeated observations of the same objects are continuously added to it. The implementation of such a database and database tools where astronomical objects extracted from survey data can be stored, updated and associated with all the information available from internal and external sources is an essential element for the successful use of the Public Survey data. The need for such databases transcends EIS and several groups in Europe involved in wide-field imaging surveys are tackling the same problem (e.g. Terapix, ΩCam). These various groups are exchanging ideas, and a preliminary implementation has recently been tested by the EIS Survey Team, in collaboration with the ESO Archive Group, using the Objectivity database. As the work on the pipeline is consolidated, the attention of the software developers in the Survey Team will focus on this critical area which is a key element for the efficient scientific exploitation of the survey data.

## 8. Conclusions

After two years of observations, software development, data reduction and distribution, nearly all the data from the ongoing Public Surveys (EIS-WIDE, EIS-DEEP and Pilot Survey) are publicly available. The only exceptions are the data gathered from observations conducted with the WFI in September 1999 and some data taken with the narrow-band *z* filter for which no proper calibration is currently available. The EIS experience has been extremely valuable for the new Public Survey to be conducted in Periods 64–67, as important lessons have been learned in handling large volumes of data, in setting up rules for automatically processing images and recipes for data quality control, and in ways and means of exporting the data to the users in an adequate form and in a timely fashion. With the accumulated experience from four previous releases and with the major development phase drawing to a close, new ways of making the data available are being considered which will hopefully provide interested users with a broader range of products as well as more flexibility and thus contribute to the scientific exploitation of the VLT.

## 9. Acknowledgments

We would like to thank A. Renzini for his effort in steering the imaging survey, the WG members for their scientific input and former EIS team members for their contribution over the years. Special thanks to E. Bertin for addressing the innumerable questions and requests posed by the EIS team and for introducing new features to SExtractor as needed and to T. Erben for his help in the analysis of the PSF.